\documentstyle[prc,aps,epsf,amssymb,amsbsy,amstext,subeqnarray,amsfonts]{revtex}
\begin{document}

\title{Chiral symmetry and quantum hadro-dynamics}

\author{G. Chanfray\protect\( ^{1}\protect  \), M. Ericson\protect\( ^{1,2}\protect  \), P.A.M. Guichon\protect\( ^{3}\protect  \)}

\address{{ \protect\( ^{1}\protect \)IPNLyon, IN2P3-CNRS et UCB
 Lyon I, F69622 Villeurbanne Cedex} }

\address{{ \protect\( ^{2}\protect \)Theory division, CERN, CH12111 Geneva} }

\address{{ \protect\( ^{3}\protect \)SPhN/DAPNIA, CEA-Saclay, F91191 Gif
sur Yvette Cedex} }

\maketitle

\begin{abstract}
Using the linear sigma model, we study the evolutions of the quark condensate
and of the nucleon mass in the nuclear medium. Our formulation of the model
allows the inclusion of both pion and scalar-isoscalar degrees of freedom. It
guarantees that the low energy theorems and the constrains of chiral perturbation
theory are respected. We show how this formalism incorporates quantum hadro-dynamics
improved by the pion loops effects. 
\end{abstract}

PACS numbers: 24.85.+p 11.30.Rd  12.40.Yx 13.75.Cs 21.30.-x

\section{Introduction }

The influence of the nuclear medium on the spontaneous breaking of chiral symmetry
remains an open problem. The amount of symmetry breaking is measured by the
quark condensate, which is the expectation value of the quark operator \( \overline{q}q \).
The vacuum value \( \left\langle \overline{q}q(0)\right\rangle  \) satisfies
the Gell-mann, Oakes and Renner relation:\begin{equation}
\label{Eq0.1}
2m_{q}\left\langle \overline{q}q(0)\right\rangle =-m_{\pi }^{2}f_{\pi }^{2},
\end{equation}
where \( m_{q} \) is the current quark mass, \( q \) the quark field, \( f_{\pi }=93MeV \)
the pion decay constant and \( m_{\pi } \) its mass. In the nuclear medium
the quark condensate decreases in magnitude. Indeed the total amount of restoration
is governed by a known quantity, the nucleon sigma commutator \( \Sigma _{N} \)
which, for any hadron \( h \) is defined as:\begin{equation}
\label{Eq2'}
\Sigma _{h}=-i\langle h| [ Q_{5},\dot{Q}_{5}] | h\rangle =2m_{q}\int d\vec{x}\left[ \left\langle h\left| \overline{q}q(\vec{x})\right| h\right\rangle -\left\langle \overline{q}q(0)\right\rangle \right] ,
\end{equation}
where \( Q_{5} \) is the axial charge and \( \dot{Q}_{5} \) its time derivative.
At low density, where the nucleons do not interact, one can estimate the restoration
effect by adding the contributions of the individual nucleons. This leads to\cite{drukarev90a,cohen92a}:
\begin{equation}
\label{Eq2''}
\frac{\left\langle \overline{q}q\left( \rho \right) \right\rangle }{\left\langle \overline{q}q\left( 0\right) \right\rangle }=1-\frac{\Sigma _{N}\, \rho }{m_{\pi }^{2}f^{2}_{\pi }},
\end{equation}
where \( \rho  \) is the density. Using the experimental value \( \Sigma _{N}\propto 50\, {\rm MeV} \)
one thus get a relative drop of almost \( 40\% \) at normal density \( \rho _{0}=0.17{\rm fm}^{-3} \). 

The part of the restoration process which is best understood arises from the
nuclear virtual pions. They act in the same way as the real ones of the heat
bath, leading to a similar expression for the evolution of the quark condensate,
that is:\begin{equation}
\label{Eq1}
\frac{\left\langle \overline{q}q\left( \rho ,T\right) \right\rangle }{\left\langle \overline{q}q\left( 0,0\right) \right\rangle }=1-\frac{\rho ^{\pi }_{s}\left( \rho ,T\right) }{2m_{\pi }f^{2}_{\pi }}=1-\frac{\left\langle \phi ^{2}\right\rangle }{2f_{\pi }^{2}},
\end{equation}
where \( T \) is the temperature and the scalar pion density \( \rho ^{\pi }_{s} \)
is linked to the average value of the squared pion field \( \phi ^{2}=\vec{\phi }.\vec{\phi } \)
through (it is understood that the vacuum contribution to \( \left\langle \phi ^{2}\right\rangle  \)
is substracted):\begin{equation}
\label{Eq2}
\rho ^{\pi }_{s}=m_{\pi }\left\langle \phi ^{2}\right\rangle .
\end{equation}
Estimates of the RHS of Eq.(\ref{Eq2}) for nuclear matter at normal density give
\( \rho _{s}^{\pi }\propto 0.07{\rm fm}^{-3} \) {\Large }which leads to a \( 20\% \)
relative decrease, due to the pion cloud, of the quark condensate. So half of
the restoration is due to the nuclear pion cloud. Concerning the manifestation
of the symmetry restoration, the pion cloud produces a correlator mixing effect
first introduced in the framework of the heat bath by Dey et al.\cite{dey90a}
and adapted to the nuclear case by Chanfray et al.\cite{chanfray98a}.

The restoration of non pionic origin is by contrast not so well understood.
In this paper we clarify the role of the meson clouds with special emphasis
on the scalar-isoscalar meson which enters in the relativistic models of nuclei.
It is expected to play a distinguished role since it has the same quantum numbers
as the condensate and can thus dissolve into it. At variance with the scalar-isoscalar
meson, which contributes already at the mean field level, the other mesons,
including the pion, contribute only through the fluctuations. On the other hand
the scalar-isoscalar meson is also an essential actor of the nuclear dynamics
since it provides the medium range attraction which binds the nucleons together
in the nucleus. In particular the existence of a scalar field is a central ingredient
of quantum hadro-dynamics (QHD)\cite{serot86}. 

The mean scalar field is responsible for the lowering of the nucleon mass (\( M^{*} \))
in the nucleus. Effective values of \( M^{*} \) lower than the free mass by
several hundreds of MeV are commonly discussed in QHD. It is quite appealing
to interpret this mass reduction as a signal for the symmetry restoration. Indeed
one scenario for the Wigner realisation of the symmetry is the vanishing of
hadronic masses. Partial restoration would then show up as a reduction of the
masses. This was the suggestion of Brown and Rho\cite{brown91a} who proposed
a scaling law linking the mass reduction to the condensate evolution according
to:\begin{equation}
\label{Eq3}
\frac{M^{*}}{M}=\frac{f^{*}_{\pi }}{f_{\pi }}=\left[ \frac{\left\langle \overline{q}q\left( \rho \right) \right\rangle }{\left\langle \overline{q}q\left( 0\right) \right\rangle }\right] ^{\alpha },\, \, \, \, {\rm with\, \, }\alpha =1/3.
\end{equation}
Birse\cite{birse96a,birse98a} pointed out the difficulties inherent to the scaling
law (\ref{Eq3}). The condensate evolution is, as we have seen previously, partly
governed by the expectation value \( \left\langle \phi ^{2}\right\rangle  \)
which contains a term of order \( m_{\pi } \). It is linked to the non analytical
part of order \( m_{\pi }^{3} \) of the pionic part of  the nucleon sigma
commutator \( \Sigma _{N} \) through the relation: \begin{equation}
\label{Eq4}
\Sigma _{N}({\rm pionic})=\frac{m_{\pi }^{2}}{2}\int d\vec{x}\left\langle N\left| \phi ^{2}(\vec{x})\right| N\right\rangle .
\end{equation}
 If the mass evolution were to follow the condensate one according to Eq.(\ref{Eq3}),
it would thus contain a term of order \( m_{\pi } \), which is forbidden by
chiral perturbation theory\cite{birse96a,birse98a}. It is thus clear that \( \left\langle \phi ^{2}\right\rangle  \),
i.e. the condensate evolution of pionic origin, cannot influence the mass.

This argument however does not imply that other actors of the restoration cannot
affect the mass. In fact the picture which naturally emerges from the previous
discussion is that different components of the restoration may produce different
signals. One of them is the axial-vector mixing induced by the pionic type of
restoration. Another signal may be the hadron mass reduction and its link to
the condensate evolution is studied in the following. 

The model we use for this study is the linear sigma model which possesses chiral
symmetry and considers the sigma and the pion as chiral partners. We first point
out that this model faces a potential problem. In the tree approximation, the
nucleon mass has its origin in the spontaneous breaking of the chiral symmetry
and is proportional to the condensate. Its in-medium value then follows the
condensate evolution. However the tree approximation is not sufficient to describe
this evolution which is largely influenced by the pion loops, as discussed previously.
If the proportionality between the mass and the condensate evolutions still
holds once pion loops are included, then the mass would be affected by the pion
loops in the same way as the condensate. But, as explained before, this is forbidden
by chiral perturbation theory.

We propose to clarify this point through a reformulation of the linear sigma
model. As is well known the predictions of the linear sigma model generally
involve cancellations between several graphs. One example is the \( \pi N \)
scattering amplitude where the contribution from sigma exchange, which by itself
violates the soft pion results, combines with the Born term to satisfy them.
We will see that this is also the case in the mass evolution problem. Namely
the mass evolution, even though it is linked to the condensate evolution, is
independent of the pion density.

To mention some previous works, Birse and McGovern\cite{birse93a} investigated
the evolution of the condensate with the density up to second order. This was
done in the usual formulation of the linear sigma model and pion loops were
not included. On the other hand Delorme et al.\cite{delorme96a} performed a
similar investigation in the non linear sigma model which is well adapted for
the pion loops but ignores the role of the scalar meson exchange. Our formulation
allows to include both effects, which is necessary in order to discuss the relation
between the mass and condensate evolutions. 

Our article is organized as follows. In Section \ref{Reminder} we remind the
steps which lead from the linear sigma model to the non linear one and we present
arguments in favor of an alternative formulation. In Section \ref{alternative},
we reformulate the linear sigma model in the standard non linear form for what
concerns the pion field but we keep explicitly a scalar degree of freedom (called
\( \theta  \)) corresponding to the fluctuation along the chiral radius. The
resulting form of the Lagrangian automatically embodies the cancellations imposed
by chiral symmetry. We tentatively identify this fluctuation with the scalar
meson which produces the nucleon nucleon attraction. In Section \ref{medium}
we discuss the prediction of this model for the behavior of various in medium
quantities and we make explicit the link with QHD. Section \ref{conclusion}
is our conclusion.

\section{Reminder of the sigma model. }
\label{Reminder}
The starting point is the usual linear sigma model\cite{gellmann58a} which
is defined by the Lagrangian:\begin{equation}
\label{Eq5}
{\cal L}=i\overline{\psi }\gamma ^{\mu }\partial _{\mu }\psi +\frac{1}{2}\left( \partial _{\mu }\sigma \partial ^{\mu }\sigma +\partial _{\mu }\vec{\pi }.\partial ^{\mu }\vec{\pi }\right) +g_{0}\overline{\psi }(\sigma +i\vec{\tau }.\vec{\pi }\gamma _{5})\psi -V_{{\rm pot}}\left( \sigma ^{2}+\pi ^{2}\right) +c\sigma ,
\end{equation}
where \( (\psi ,\sigma ,\vec{\pi }) \) are respectively the nucleon, sigma
and pion fields, the arrow indicating the isovector character of the pion. For
the meson potential \( V_{{\rm pot}} \) we take the usual form \( V_{{\rm pot}}(x)=\lambda (x-v^{2})^{2}/4. \)

For later use it is convenient to write \( {\cal L} \) in terms of the \( 2\times 2 \)
matrix \( W=\sigma +i\vec{\tau }.\vec{\pi } \) acting in the nucleon isospin
space. Noting \( P^{R/L}=(1\pm \gamma _5)/2 \) the chirality projectors,
one can write\begin{eqnarray}
{\cal L} & = & {\cal L}_{0}+{\mathcal{L}}_{\chi SB}\label{Eq6a} \\
{\cal L}_{0} & = & i\overline{\psi }\gamma ^{\mu }\partial _{\mu }\psi +g_{0}\overline{\psi }\left[ WP^{R}+W^{\dagger }P^{L}\right] \psi +\nonumber \label{Eq15} \\
 &  & \frac{1}{4}Tr\, \partial _{\mu }W\, \partial ^{\mu }W^{\dagger }-\frac{\lambda }{4}\left( \frac{1}{2}Tr\, WW^{\dagger }-v^{2}\right) ^{2},\label{Eq6b} \\
{\mathcal{L}}_{\chi SB} & = & c\sigma =\frac{c}{2}Tr\, W.\label{Eq6c} 
\end{eqnarray}
In this form it is apparent that \( {\cal L}_{0} \) is invariant under the
transformations \[
P^{R}\psi \rightarrow g_{R}\, P^{R}\psi ,\, \, \, \, P^{L}\psi \rightarrow g_{L}\, P^{L}\psi ,\, \, \, \, W\rightarrow g_{L}Wg_{R}^{\dagger }\]
where \( (g_{R},\, g_{L}) \) are elements of the \( SU(2)\times SU(2) \) group.
The term \( {\mathcal{L}}_{\chi SB} \) breaks explicitly the symmetry.

In the vacuum one has \( \left\langle \vec{\pi }\right\rangle =0 \) by parity
and one notes \( \left\langle \sigma \right\rangle =f_{\pi } \) the constant
expectation value of \( \sigma . \) The breaking of the symmetry by the vacuum
(\( \left\langle \sigma \right\rangle \neq 0) \) is realized at the classical
level by imposing that the mesons energy be stationary at the point \( (\sigma =f_{\pi },\, \vec{\pi }=0) \).
This amounts to \begin{equation}
\label{Eq6}
\frac{\partial }{\partial \sigma }\left[ V_{{\rm pot}}\left( \sigma ^{2}+\pi ^{2}\right) -c\sigma \right] _{\sigma =f_{\pi },\, \vec{\pi }=0}=0,
\end{equation}
since the stationarity with respect to \( \vec{\pi } \) is trivially satisfied.
The other parameters are fixed by identifying the mass terms, that is \begin{equation}
\label{Eq7}
g_{0}\left\langle \sigma \right\rangle =M_{N},\, \, \, \, m_{\sigma }^{2}=\left. \frac{\partial ^{2}V_{{\rm pot}}}{\partial \sigma ^{2}}\right| _{\sigma =f_{\pi },\, \vec{\pi }=0},\, \, \, \, m_{\pi }^{2}\, \delta (i,j)=\left. \frac{\partial ^{2}V_{{\rm pot}}}{\partial \pi _{i}\partial \pi _{j}}\right| _{\sigma =f_{\pi },\, \vec{\pi }=0}.
\end{equation}
One gets\begin{equation}
\label{Eq9}
c=m_{\pi }^{2}\, f_{\pi },\, \, \, \, \lambda =\frac{m_{\sigma }^{2}-m_{\pi }^{2}}{2f_{\pi }^{2}},\, \, \, \, v^{2}=f_{\pi }^{2}\frac{m_{\sigma }^{2}-3m_{\pi }^{2}}{m_{\sigma }^{2}-m_{\pi }^{2}},
\end{equation}
and the quantized version of the model is obtained by considering \( \vec{\pi } \)
and \( \sigma '=\sigma -f_{\pi } \) as the degrees of freedom.

This model is referred to as the linear sigma model (LSM). Since, in the limit
\( m_{\pi }\rightarrow 0 \), its equations of motion respect chiral symmetry
this model reproduces the soft pion theorems in the tree approximation. However
this generally involves somewhat unnatural cancellations between several diagrams.
Moreover the lack of experimental evidence (see however Ref.\cite{tornqvist96a})
for a scalar meson that could be associated with the fluctuation \( \sigma ' \)
has led to the idea that this field was unphysical and should be eliminated
from the model. This is achieved by letting \( m_{\sigma }\rightarrow \infty  \)
which leads to the constraint\begin{equation}
\label{Eq10}
\sigma ^{2}+\pi ^{2}=f_{\pi }^{2}
\end{equation}
 for the finite energy solutions. The constraint (\ref{Eq10}) is solved by
the point transformation\begin{equation}
\label{Eq11}
\sigma =f_{\pi }\cos F\left( \frac{\phi }{f_{\pi }}\right) ,\, \, \, \, \vec{\pi }=\hat{\phi }\sin F\left( \frac{\phi }{f_{\pi }}\right) 
\end{equation}
 which eliminates the \( \sigma  \) field and defines \( \vec{\phi }=\phi \hat{\phi } \)
as a new pion field. \( F \) is an odd function of the form\begin{equation}
\label{Eq12}
F(x)=x+\alpha x^{3}+\cdots 
\end{equation}
which selects the particular realisation of the model. Changing \( F \) amounts
to a redefinition of the pion field and thus should not affect the physics.
In the following we keep \( \alpha  \) arbitrary and check that the final results
do not depend on it. 

The last step is to perform a new point transformation defined by\cite{weinberg68a}:\begin{equation}
\label{Eq13}
\psi =\exp \left( -\frac{1}{2}i\vec{\tau }.\hat{\phi }F(\phi /f_{\pi })\gamma _{5}\right) N
\end{equation}
 and to take \( N \) as the nucleon field. This defines the Non Linear Sigma
Model. Due to the transformation (\ref{Eq13}), the pion then couples to the
nucleon \( N \) only through derivatives. This eliminates the unnatural cancellations
of the LSM because the Born terms are automatically suppressed by powers of
\( m_{\pi } \) in the soft pion limit.

This is all fine for chiral symmetry but somewhat frustrating for nuclear physics.
The reason is that the medium range attraction is known to be dominated by a
scalar-isoscalar correlated 2-pion exchange. Chiral perturbation theory actually
forbids the identification of this attraction with the exchange of the \( \sigma  \)'
(\( =\sigma -f_{\pi } \) ) field but, if we go back in the above discussion,
we realize that the chiral radius \( \sqrt{\sigma ^{2}+\pi ^{2}} \) has been
fixed to \( f_{\pi } \) by mere convenience. Nothing prevent us from keeping
it as a degree of freedom and to, tentatively, identify it with the meson which
produces the medium range attraction. To avoid any confusion with \( \sigma  \),
the chiral partner of the pion, we shall note it \( \Theta . \) The fact that
no such meson is clearly seen in \( \pi \pi  \) scattering is not an obstacle.
There is in the model a strong \( \Theta \rightarrow \pi \, \pi  \) coupling
which, as in the linear sigma model, leads to a large \( \Theta  \) width.
This may explain why this meson is so elusive. For the \( NN \) interaction
this large on-shell width of the \( \Theta  \) is not a conceptual difficulty
because it comes into play only through space-like exchange between nucleons.
So its width is effectively zero. 

We stress that, with respect to the LSM, we simply make a convenient change
of variables \( (\sigma ,\vec{\pi })\rightarrow (\Theta ,\vec{\phi }) \) which
avoids keeping track of the cancellations inherent to the model. When studying
elementary processes these cancellations are just a matter of care, but when
they are intertwined with the unavoidable approximations of the nuclear many
body problem this may lead to results inconsistent with chiral symmetry.

\section{Alternative formulation of the Linear Sigma Model}
\label{alternative}
Our starting point is defined by the Lagrange density:\begin{eqnarray}
{\cal L}={\cal L}_{0}+\Delta {\cal L}_{0} & +{\mathcal{L}}_{\chi SB},\label{Eq14} 
\end{eqnarray}
 where to the symmetric piece \( {\cal L}_{0} \) defined in Eq.(\ref{Eq6b})
we have added, as in Ref.\cite{chanfray98a}, the chiral invariant piece: \begin{equation}
\label{Eq16}
\Delta {\cal L}_{0}=ia\overline{\psi }\gamma ^{\mu }\, \left( W\partial _{\mu }W^{\dagger }\, P^{L}+W^{\dagger }\partial _{\mu }W\, P^{R}\right) \psi ,
\end{equation}
 which is not present in the original sigma model. Its only role is to generate
an axial coupling constant \( g_{A} \) different from unity in the tree approximation.
The spirit of this is not to try to make a realistic description of the nucleon
but to make easier the identification of the evolution of this quantity. The
axial current corresponding to (\ref{Eq14}) is:\begin{eqnarray}
\vec{J}_{5\mu } & = & \left( 1-\frac{a}{2}TrW^{\dagger }W\right) \left( \overline{\psi }\gamma _{\mu }\gamma _{5}\frac{\vec{\tau }}{2}\psi \right) -a\overline{\psi }\gamma _{\mu }\left( W^{\dagger }\frac{\vec{\tau }}{2}WP_{R}-W\frac{\vec{\tau }}{2}W^{\dagger }P_{L}\right) \psi \nonumber \\
 &  & -\frac{i}{4}Tr\left( W^{\dagger }\partial _{\mu }W+W\partial _{\mu }W^{\dagger }\right) ,\label{Eq16bis} 
\end{eqnarray}
from which one sees that one needs\begin{equation}
\label{Eq17}
a=\frac{1-g_{A}}{2f_{\pi }^{2}},
\end{equation}
 in order to get the correct value of the nucleon axial charge in the tree approximation.
 The other parameters \( (\lambda, \, \, v,\, g_{0},\, c) \) have the same expressions
as in Eqs(\ref{Eq7},\ref{Eq9}). \textbf{}Notice that if we note \( Q_{i}^{5} \)
the axial charge of the model, the symmetry breaking part of \( {\cal L} \)
is such that the identity \( H_{\chi SB}=[Q_{i}^{5},[Q^{5}_{i},H]] \) is satisfied,
as in QCD itself. 

Guided by the discussion of Section \ref{Reminder} we make the point transformation
\( (\sigma ,\vec{\pi })\rightarrow (\Theta ,\vec{\phi }) \) defined by:\begin{equation}
\label{Eq18}
\sigma =\Theta \cos F\left( \frac{\phi }{f_{\pi }}\right) ,\, \, \, \, \vec{\pi }=\Theta \, \hat{\phi }\sin F\left( \frac{\phi }{f_{\pi }}\right) ,
\end{equation}
which allows to write:\begin{equation}
\label{Eq19}
W=\Theta \, U,\, \, \, \, U=\xi ^{2}=\exp \left( i\vec{\tau }.\hat{\phi }F\left( \frac{\phi }{f_{\pi }}\right) \right) ,
\end{equation}
 and we define the new nucleon field: \begin{equation}
\label{Eq20}
N=\left( \xi \, P^{R}+\xi ^{\dagger }\, P^{L}\right) \psi ,
\end{equation}
which is equivalent to Eq.(\ref{Eq13}). Note that the mass term \( \overline{N}N \)
is a chiral invariant. In the vacuum one has \( \left\langle \Theta \right\rangle =\left\langle \sigma \right\rangle =f_{\pi } \)
. So we define the fluctuation \( \theta =\Theta -f_{\pi } \) and write \( {\cal L} \)
in terms of the degrees of freedom \( (N,\theta ,\vec{\phi }) \), that is: 

\begin{eqnarray}
{\mathcal{L}} & = & \left( f_{\pi }+\theta \right) ^{2}\, Tr\partial ^{\mu }U\partial _{\mu }U^{\dagger }\, +\, {1\over 2}\partial ^{\mu }\theta \partial _{\mu }\theta \, -\, {m^{2}_{\sigma }-m^{2}_{\pi }\over 8f^{2}_{\pi }}\left( \theta ^{2}+2f_{\pi }\theta \, +\, {2f^{2}_{\pi }m^{2}_{\pi }\over m^{2}_{\sigma }-m^{2}_{\pi }}\right) ^{2}\nonumber \\
 &  & +i\bar{N}\gamma ^{\mu }\partial _{\mu }N\, -\, M_{N}\left( 1+{\theta \over f_{\pi }}\right) \bar{N}N\nonumber \\
 &  & +\bar{N}\gamma _{\mu }{\mathcal{V}}^{\mu }_{c}N\, +\left( 1\, -(1-g_{A})\left( 1+{\theta \over f_{\pi }}\right) ^{2}\right) \, \bar{N}\gamma _{\mu }\gamma ^{5}{\mathcal{A}}^{\mu }_{c}N\nonumber \\
 &  & +i{1-g_{A}\over 2f_{\pi }}\left( 1+{\theta \over f_{\pi }}\right) \bar{N}\gamma ^{\mu }N\, \partial _{\mu }\theta +{\mathcal{L}}_{\chi SB},\label{Eq21} 
\end{eqnarray}
 where we have defined: \begin{equation}
\label{Eq22}
{\mathcal{V}}^{\mu }_{c}={i\over 2}\left( \xi \partial _{\mu }\xi ^{\dagger }+\xi ^{\dagger }\partial _{\mu }\xi \right) \, \qquad {\mathcal{A}}^{\mu }_{c}={i\over 2}\left( \xi \partial _{\mu }\xi ^{\dagger }-\xi ^{\dagger }\partial _{\mu }\xi \right) .
\end{equation}
We have:\begin{equation}
\label{Eq22'}
m_{\theta }^{2}=\left. \frac{\partial ^{2}{\mathcal{L}}(\theta )}{\partial \theta ^{2}}\right| _{\theta =0}=m_{\sigma }^{2},
\end{equation}
so in the following \( m_{\sigma } \) will be replaced by \( m_{\theta } \). 

In terms of the new variables, we get the following expressions for the symmetry
breaking piece: \begin{equation}
\label{Eq23}
{\mathcal{L}}_{\chi SB}=f_{\pi }m^{2}_{\pi }\, \big (f_{\pi }+\theta \big )cosF\left( \frac{\phi }{f_{\pi }}\right) ,
\end{equation}
 and for the axial current:\begin{eqnarray}
\vec{J}_{5\mu } & = & -i{f^{2}_{\pi }\over 4}\left( 1+{\theta /f_{\pi }}\right) ^{2}\, Tr\left( \vec{\tau }U^{\dagger }\partial _{\mu }U\, -\vec{\tau }U\partial _{\mu }U^{\dagger }\right) \nonumber \\
 &  & +{1\over 2}\left( 1\, -(1-g_{A})\left( 1+{\theta /f_{\pi }}\right) ^{2}\right) \, \bar{N}\gamma _{\mu }\gamma ^{5}\left( \xi {\tau _{i}\over 2}\xi ^{\dagger }\, +\, \xi ^{\dagger }{\tau _{i}\over 2}\xi \right) N\nonumber \\
 &  & +{1\over 2}\bar{N}\gamma _{\mu }\left( \xi {\tau _{i}\over 2}\xi ^{\dagger }\, -\, \xi ^{\dagger }{\tau _{i}\over 2}\xi \right) N.\label{Eq24} 
\end{eqnarray}

Some comments on the Lagrangian of Eq.(\ref{Eq21} ) are in order. The term
\( \bar{N}\gamma _{\mu }\gamma ^{5}{\mathcal{A}}^{\mu }_{c}N \) generates the
standard \( \pi NN \) p-wave coupling but corrected by a \( 3\pi NN \) coupling
and other higher order terms. One can check that the Goldberger relation \( g_{\pi NN}\, f_{\pi }=M_{N}\, g_{A} \)
is fulfilled. There is a non derivative \( \theta N \) interaction with a coupling
constant equal to \( M_{N}/f_{\pi }\propto 10 \) which is smaller than the
\( \pi N \) coupling constant by a factor \( 1/g_{A}. \) Note that this coupling
constant is not a free parameter in this model because all the nucleon mass
is generated by the spontaneous symmetry breaking. This will no longer be true
in models where part of the nucleon mass is due to the confinement. Finally
we stress that, in the chiral limit, the new scalar field \( \theta  \) couples
only derivatively to two pions, there is no term of the form \( \theta \, \phi ^{2} \).
This insures the validity of the soft pion theorems for \( \pi N \) scattering.

\section{Medium effects }
\label{medium}
We are now in a situation to describe various in-medium quantities in the framework
of the mean field approximation combined with the pion gas limit.

\subsection{Condensate evolution}

Firstly the quark condensate and its evolution at finite density can be obtained
by identifying the symmetry breaking pieces of QCD and the one of our Lagrangian,
that is \begin{equation}
\label{Eq25}
-2m_{q}\bar{q}q\, \leftrightarrow \, f_{\pi }m_{\pi }^{2}\sigma =f_{\pi }m^{2}_{\pi }\, \left( f_{\pi }+\theta \right) \, \cos F\left( \frac{\phi }{f_{\pi }}\right) .
\end{equation}
 This equation shows that the condensate evolution is driven by the mean value
of \( \sigma , \) the chiral partner of the pion. From Eq.(\ref{Eq25}) and
using the Gellmann, Oakes and Renner relation, we get the relative modification
of the condensate: \begin{equation}
\label{Eq26}
\frac{\left\langle \overline{q}q(\rho )\right\rangle }{\left\langle \overline{q}q(0)\right\rangle }=1-\frac{\left\langle \phi ^{2}\right\rangle }{2f_{\pi }^{2}}+\langle \frac{\theta }{f_{\pi }}\left( 1-\frac{\phi ^{2}}{2f_{\pi }^{2}}\right) \rangle +\cdots 
\end{equation}
 where we have expanded \( \cos F(\phi /f_{\pi }) \) and kept only the leading
terms in \( 1/f_{\pi } \). There are two contributions to the restoration effect.
The first one arises from the pion cloud, the second one is driven by the scalar
field \( \theta  \). The second contribution also depends on the squared pion
field, which is to be expected since the condensate is not a chiral invariant
quantity.

The mean field \( \left\langle \theta \right\rangle  \) is obtained from the
equation of motion which writes, for a uniform medium of density \( \rho  \):\begin{equation}
\label{Eq27}
\frac{f_{\pi }m_{\theta }^{2}}{2}\left( 2X+3X^{2}+X^{3}\right) +g_{0}\rho =0,\, \, \, \, \, {\rm with}\, \, \, X=\frac{\left\langle \theta \right\rangle }{f_{\pi }},
\end{equation}
 where terms of order \( m_{\pi }^{2} \) have been ignored. To second order
in the density the solution is: \textbf{\Large }\begin{equation}
\label{Eq28}
\frac{\left\langle \theta \right\rangle }{f_{\pi }}=-\frac{g_{0}\rho }{f_{\pi }m_{\theta }^{2}}-\frac{3}{2}\left( \frac{g_{0}\rho }{f_{\pi }m_{\theta }^{2}}\right) ^{2}.
\end{equation}
 To zeroth order in the pion field the quantity \( \left\langle \theta \right\rangle /f_{\pi } \)
fixes the relative amount of restoration from the \( \theta  \) exchange. Numerical
estimates will be discussed later. 

\begin{figure}
\epsfysize=6cm
\centerline{\epsfbox{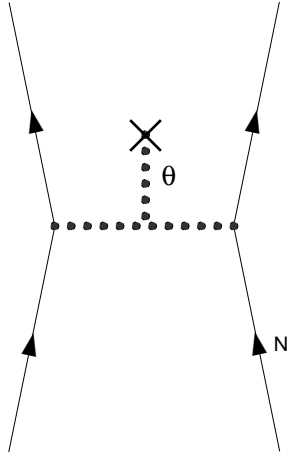}}
\caption{\label{Fig-theta-exch}The mechanism which induces the term quadratic in \protect\( \rho \protect \)
in Eq.(\ref{Eq28})}
\end{figure}

The term quadratic in the density in Eq.(\ref{Eq28}) represents a moderate
effect at normal density but its interpretation is interesting. Due to the \( \theta ^{3} \)
vertex in Eq.(\ref{Eq21}), the mean field \( \left\langle \theta \right\rangle  \)
gets a contribution from the \( \theta  \) exchange as shown on Fig.\ref{Fig-theta-exch}.
For what concerns the condensate evolution this second order term in \( \rho  \)
represents the contribution to the restoration due to the \( \theta  \) exchanged
between nucleons. Indeed the sigma commutator of the \( \theta  \) meson, \( \Sigma _{\theta } \),
which fixes the amount of restoration induced by a single \( \theta  \), can
be obtained from the Feynman-Hellman theorem, which leads to:

\begin{equation}
\label{Eq29}
\Sigma _{\theta }=c\left. \frac{dm_{\theta }}{dc}\right| _{\lambda ,v}=3c\frac{m_{\pi }}{m_{\theta }}\left. \frac{dm_{\pi }}{dc}\right| _{\lambda ,v}=3\frac{m_{\pi }}{m_{\theta }}\Sigma _{\pi }=\frac{3m_{\pi }^{2}}{2m_{\theta }}.
\end{equation}
This quantity  has to be multiplied by the two-body contribution to the \( \theta  \)
scalar density \( \rho ^{(2)}_{\theta } \) which is:\begin{equation}
\label{Eq29'}
\rho ^{(2)}_{\theta }=m_{\theta }\, \int d\vec{x}_{1}d\vec{x}_{2}\, \rho (\vec{x}_{1})\rho (\vec{x}_{2})\theta _{1}(\vec{x})\theta _{2}(\vec{x})=\frac{g_{0}^{2}\rho ^{2}}{m_{\theta }^{3}},
\end{equation}
 where \( \theta _{i} \) is the \( \theta  \) field created by the nucleon
located at \( \vec{x}_{i} \). {\Large }In order to obtain the relative amount
of restoration we have to multiply Eq.(\ref{Eq29}) by expression (\ref{Eq29'})
and divide by \( (-f_{\pi }^{2}m_{\pi }^{2}). \) This gives \( 3(g_{0}\rho /m\sigma ^{2})^{2}/2 \),
which is precisely the quadratic part of Eq.(\ref{Eq28}). This second order
term in the density of the quark condensate was already given by Birse and McGovern\cite{birse93a}.
It is absent in the non linear sigma model\cite{delorme96a}, as it should be
since this correction concerns only the \( \theta  \) meson. Weise\cite{weise96a}
also estimated the contribution to the restoration of the sigma exchanged between
nucleons by relating the sigma commutator of the sigma to the nucleon one.

\subsection{Nucleon and \protect\( \theta \protect \) mass evolution}

We now come to another in-medium quantity, the effective nucleon mass. Its evolution
in presence of the mean scalar field is apparent from the Lagrangian (\ref{Eq14}).
It reads : \begin{equation}
\label{Eq30}
M_{N}^{*}=M_{N}-g_{0}\left\langle \theta \right\rangle =M_{N}-\frac{g_{0}^{2}\rho }{m_{\theta }^{2}}-\frac{3g_{0}}{2f_{\pi }}\left( \frac{g_{0}^{2}\rho }{m_{\theta }^{2}}\right) ^{2}.
\end{equation}
It is exclusively governed by the chiral invariant scalar field \( \theta  \).
It has no dependence at all on \( \phi ^{2} \), which eliminates the conflict
with the chiral perturbation constraints. The cancellations of the linear model
are indeed present for the mass, in such a way that the influence of the pion
loops on the mass is eliminated. The present formulation of the model automatically
insures these cancellations. The identification with the mean scalar field of
the Walecka model now becomes obvious. The scalar field of QHD has to be identified
with the scalar invariant mean field \( \left\langle \theta \right\rangle  \)
and not with \( \sigma '=\sigma -f_{\pi } \), the chiral partner of the pion.
The model also provides its coupling to the nucleon, \( g_{\theta NN}=g_{0}=M/f_{\pi }\simeq 10, \)
somewhat smaller than \( g_{\pi nN}=13.5, \) a welcome feature with respect
to the phenomenology of QHD\cite{serot86}. 

The chiral invariant character of the mean scalar field of QHD is not a new
concept. In the work of Serot\cite{furnstahl97a} a chiral invariant scalar
is added to the non linear sigma model but its coupling to the nucleon is arbitrary
and the in-medium mass bears no relation to the condensate evolution. On the
other hand Delorme et al.\cite{delorme00a} have studied the nucleon mass in
the quark-meson coupling model\cite{guichon88a}. Here the source of the scalar
field is the quark scalar density, which allows a link between the nucleon mass
and the condensate modifications. The chiral invariant character of the quark-meson
coupling was imposed.

We are now in a situation to discuss the relation between the mass and condensate
evolutions, a connection totally absent in the standard formulation of QHD,
and the connection between the present work and the scaling law of Brown and
Rho. The mass evolution is related to the condensate evolution, but only to
part of it, the purely non-pionic part. Only the chiral invariant field \( \theta  \)
influences the mass. This \( \theta  \) field is dressed by the pion loops
while the \( \sigma  \) is not. It is only to zero order in the pion loops
that the two evolutions are the same. Note that, to zeroth order in the pion
loops and in the low density limit, our approach gives \( \alpha =1 \) instead
of \( \alpha =1/3 \) for the expression of the scaling factor of Eq.(\ref{Eq3}). 

In the same way the effective \( \theta  \) mass follows from the Lagrangian
(\ref{Eq21}). In the nuclear medium the \( \theta  \) field acquires mean
value \( \left\langle \theta \right\rangle  \) and the effective mass refers
to the fluctuations about this mean value, that is:\begin{equation}
\label{Eq31'}
m_{\theta }^{*2}=\left. \frac{\partial ^{2}{\mathcal{L}}(\theta )}{\partial \theta ^{2}}\right| _{\theta =\left\langle \theta \right\rangle }
\end{equation}
 which leads to\begin{equation}
\label{Eq31}
\left( \frac{m_{\theta }^{*}}{m_{\theta }}\right) ^{2}=1+3\frac{\left\langle \theta \right\rangle }{f_{\pi }}+\frac{3\left\langle \theta \right\rangle ^{2}}{2f_{\pi }^{2}}=1-\frac{3g_{0}\rho }{f_{\pi }m_{\theta }^{2}}.
\end{equation}
 It turns out that in Eq.(\ref{Eq31}) there is a cancellation between the terms
which are quadratic in \( \rho  \). The \( \theta  \) mass is reduced by the
medium effects getting closer to the pion mass. It follows a pattern similar
to that of the nucleon mass, with a somewhat faster evolution, as seen by comparing
Eqs.(\ref{Eq30}, \ref{Eq31}). Thus in the nuclear medium, the shape of the
Mexican hat (\( V_{{\rm pot}}) \) is appreciably modified. There is not only
a shrinking of the radius of the chiral valley due to the mean value of the
\( \theta  \) field, but accordingly the potential becomes more shallow. The
lowering of the \( \theta  \) mass suggests enhanced fluctuations around the
mean value. The connection between chiral symmetry restoration and the sigma
mass as well as the experimental implications have  been studied by Hatsuda
et al.\cite{hatsuda99a}

We now make some numerical evaluations. We have five paramaters in our version
of the model. They are linked, through the set of Eqs.(\ref{Eq7}, \ref{Eq9}),
to the pion decay constant, the nucleon mass, the axial coupling constant and
the pion and the theta masses. All these quantities are measured, but the \( \theta  \)
mass which can be taken as a free parameter. As an example we will take two
values: \( m_{\theta }=1GeV \) and \( m_{\theta }=0.8GeV \) . For \( m_{\theta }=1GeV \)
(resp. \( 0.8GeV \) ) the scalar mean field \( \theta  \) has a value of \( 17MeV \)
, (resp. \( 30MeV \)) at normal density. The corresponding nucleon mass reduction
are: \( M^{*}_{N}-M_{N}=-170MeV \) (resp. \( -300MeV \)). These magnitudes
are compatible with the current phenomenology of QHD. According to Eq.(\ref{Eq31}),
the \( \theta  \) mass also drops by \( 22\% \), (resp. \( 34\% \)), an appreciable
modification\textbf{.} For the condensate evolution we remind that the pion
cloud, that is the term \( \left\langle \phi ^{2}\right\rangle /2f_{\pi }^{2} \)
in Eq.(\ref{Eq26}), produces a relative decrease of about 20\%. The part of
the condensate evolution due to the scalar field depends not only on the expectation
value \( \left\langle \theta /f_{\pi }\right\rangle  \) but also on \( \left\langle \left( \theta /f_{\pi }\right) \left( \phi ^{2}/2f_{\pi }^{2}\right) \right\rangle  \)
which we estimate as \( \left\langle \theta /f_{\pi }\right\rangle \left\langle \phi ^{2}/2f_{\pi }^{2}\right\rangle  \).
In this way we find a relative decrease of 14\% (resp. 25\%) of the condensate.

\subsection{Evolution of \protect\( g_{A}\protect \)}

To order \( \phi ^{3} \) the axial current (\ref{Eq24}) writes\begin{eqnarray}
\vec{J}^{\mu 5} & = & f_{\pi }\left( 1+\frac{2\theta }{f_{\pi }}\right) \left[ \left( 1+3\alpha \frac{\phi ^{2}}{f_{\pi }^{2}}\right) \partial ^{\mu }\vec{\phi }+\frac{1}{f_{\pi }^{2}}\left( 2\alpha +\frac{2}{3}\right) \left( \vec{\phi }\, \vec{\phi }.\partial ^{\mu }\vec{\phi }-\phi ^{2}\partial ^{\mu }\vec{\phi }\right) \right] \nonumber  \\
 & + & g_{A}\left( 1+2\frac{g_{A}-1}{g_{A}}\frac{\theta }{f_{\pi }}\right) \left[ \overline{N}\gamma ^{\mu }\gamma ^{5}\frac{\vec{\tau }}{2}N+\frac{1}{2f_{\pi }^{2}}\overline{N}\gamma ^{\mu }\gamma ^{5}\left( \vec{\phi }\frac{\vec{\phi }.\vec{\tau }}{2}-\phi ^{2}\frac{\vec{\tau }}{2}\right) N\right] \nonumber \\
 & + & \frac{1}{f_{\pi }}\overline{N}\gamma ^{\mu }\frac{\vec{\phi }\times \vec{\tau }}{2}N+\cdots \label{A24} 
\end{eqnarray}
The last term in Eq.(\ref{A24}) does not contribute in the mean field approximation.
\textbf{}The medium modification is due to the coupling to the \( \theta  \)
field and to the terms with several pion fields. In the mean field approximation
we replace \( \theta  \) by \( <\theta > \) and \( \phi ^{i}\phi ^{j} \)
by \textbf{\( <\phi ^{i}\phi ^{j}>=<\phi ^{2}>\delta _{ij}/3 \)} which gives
the mean current

\begin{eqnarray}
<\vec{J}^{\mu 5}> & = & f_{\pi }\left( 1+\frac{2<\theta >}{f_{\pi }}\right) \left[ 1+\frac{<\phi ^{2}>}{f_{\pi }^{2}}\left( \frac{5\alpha }{3}-\frac{4}{9}\right) \right] \partial ^{\mu }\vec{\phi }\nonumber  \\
 & + & g_{A}\left( 1+2\frac{g_{A}-1}{g_{A}}\frac{<\theta >}{f_{\pi }}\right) \left( 1-\frac{<\phi ^{2}>}{3f_{\pi }^{2}}\right) \overline{N}\gamma ^{\mu }\gamma ^{5}\frac{\vec{\tau }}{2}N,\label{A25} 
\end{eqnarray}
from which, to lowest order in \( <\theta > \) and \( <\phi ^{2}>, \) we get
the following expression for the evolution of \( g_{A} \): \begin{eqnarray}
\frac{g_{A}^{*}}{g_{A}} & = & \left( 1+2\frac{g_{A}-1}{g_{A}}\frac{<\theta >}{f_{\pi }}-\frac{2}{3}\frac{<\phi ^{2}>}{2f_{\pi }^{2}}\right) \label{A27a} 
\end{eqnarray}
 At normal nuclear density the scalar contribution yields a quenching of \( g_{A} \)
of the order of \( 6\% \) while the pionic contribution gives a quenching of
about \( 15\% \). As pointed out in Ref.\cite{chanfray98a}, this result is
strictly valid only when the short range correlations between nucleons are neglected.
The renormalizations of the weak coupling constants \( g_{A} \) and \( f_{\pi } \)
, due to the suppression of the quark condensate by the scalar meson have been
previously discussed by Akmedov\cite{akmedov89a} in the usual formulation of
the sigma model, ignoring the pion loops.

\subsection{Pionic properties evolution}

We now turn to the in-medium values of pionic properties: the pion decay constant
and the pion mass. We stress that we are concerned only by the influence of
the two mesons present in our model, \( \theta  \) and \( \pi . \) In this
context the nucleons act only as a source for these fields. The problem reduces
to the question of the pion mass and decay constant in a pion gas and in the
presence of a mean scalar field \( <\theta > \)\( . \) Within this limited
framework we do not expect a realistic description of the in-medium effects
for these two quantities. Indeed the pion mass modification is linked, in the
dilute gas limit, to the isospin symmetric \( \pi N \) amplitude and is of
order \( m_{\pi }^{2}. \) It is subject to other influences than just the pion
and the theta. The pion decay constant which is linked to the pion mass by the
Gell-mann, Oakes and Renner relation is also subject to these extra influences.
Therefore we quote the implications of the model for \( f_{\pi } \) and \( m_{\pi } \)
only to show the absence of a universal link between their evolution and the
condensate one.

For what concerns the influence of the nuclear pion gas, it has already been
studied\cite{chanfray98a,chanfray99a}. It is described through the scalar pion
density (\ref{Eq2}). On this particular point the present work brings nothing
new. The novel part concerns the influence of the \( \theta  \) field. For
completeness however we treat the two effects simultaneously in our formulation
of the linear sigma model.

The effective pion decay constant is the coefficient of \( \partial ^{\mu }\vec{\phi } \)
in the mean axial current (\ref{A25}) multiplied by the wave function renormalization
\( \sqrt{Z} \) (see Appendix). To leading order in \( <\theta > \) and \( <\phi ^{2}> \)
we get the result, independent of \( \alpha  \) as it should be:\begin{eqnarray}
\frac{f_{\pi }^{*}}{f_{\pi }} & = & \left( 1+\frac{<\theta >}{f_{\pi }}-\frac{2}{3}\frac{<\phi ^{2}>}{2f_{\pi }^{2}}\right) .\label{A26a} 
\end{eqnarray}
We see that the evolution of \( f_{\pi } \) follows the condensate, Eq.(\ref{Eq26}),
only for what concerns the scalar field piece. At variance with the nucleon
mass case there is a pionic piece. With respect to the condensate evolution
(\ref{Eq26}) this pionic term is multiplied by 2/3, exactly as in the thermal
case.

The effective pion mass is defined as the position of the pole energy of the
propagator for vanishing 3-momentum. It obeys the relation\[
m_{\pi }^{*2}=m_{\pi }^{2}+S(m_{\pi }^{*},\vec{0}),\]
 where \( S(q) \) is the pion self-energy. As shown in the Appendix this leads
to \begin{eqnarray}
\left( \frac{m_{\pi }^{*}}{m_{\pi }}\right) ^{2} & = & \left( 1-\frac{<\theta >}{f_{\pi }}+\frac{<\phi ^{2}>}{6f_{\pi }^{2}}\right) .\label{A28} 
\end{eqnarray}
 Both terms on the RHS of Eq.(\ref{A28}) are positive, corresponding to a repulsive
interaction and it is clear that the evolution of \( m_{\pi } \) is completly
different from the condensate one.

\section{Conclusion }
\label{conclusion}
In this work we have studied the role of the scalar meson both in the partial
restoration of chiral symmetry and in the lowering of the hadron mass in the
nuclear medium, as well as the link between the two effects in the framework
of the linear sigma model. We have used a formulation of the linear sigma model
with the usual non linear realisation for the pion field but we have kept a
scalar degree of freedom corresponding to the fluctuation along the chiral radius.
This new scalar field \emph{is not} the chiral partner of the pion but instead
is a chiral invariant. It is already dressed by the pion loops. Its mean value
in the medium represents the modification of the radius of the chiral circle
as compared to the vacuum value. In this formalism the low energy theorems and
the constraints of chiral perturbation theory are easily fulfilled without need
for cancellations. For instance this scalar \( \theta  \) couples derivatively
to two pions, in the chiral limit. For what concerns the density evolution of
the quark condensate, which is not a chiral invariant quantity, it is instead
governed by the mean sigma field, the chiral partner of the pion. This difference
shows up in the comparison between the two evolutions. The condensate is influenced
by the pion cloud while the mass is not. It is only in to zero order in the
pion loops that the two relative evolutions become the same. In practice the
difference is large since about half of the restoration originates from the
pion cloud. 

Our work shows that for what concerns the scalar field, quantum hadro-dynamics
can be incorporated in a chiral theory such as the linear sigma model. The
scalar 
field of QHD should be identified with  the chiral invariant
scalar field \( \theta . \) This is in fact imposed by the constraints of chiral
perturbation theory, which prevents the nucleon-nucleon potential to be influenced
by the pion density in the chiral limit. Our formulation is more complete
than that of QHD in the sense that it incorporates the effect of the pion loops
in a way consistent with the constrains of chiral perturbation theory. The phenomenology
which comes out from this reformulation is compatible with that of QHD.

Concerning the signals associated with the restoration they are of two types
depending on the origin of the restoration. In the linear sigma model the quark
condensate depends on the average sigma field. In the vacuum \( \left\langle \sigma \right\rangle =f_{\pi }. \)
In the medium this quantity is modified by two effects. On the one hand there
is an oscillation along the chiral circle induced by the pionic fluctuations,
which is described by pion density. On the other hand there is a modification
of the radius of the chiral circle due to the mean scalar field \( \theta . \)
The oscillation along the chiral circle shows up in the mixing of the axial
and vector correlators. The shrinking of the chiral radius shows up in the lowering
of the nucleon mass. Both types of signal are simultaneously present in the
nuclear medium.

Acknowledgments: We have benefited from discussions with M. Birse. This work
was initiated during the stay of two of us (M.E. and P.A.M.G.) at the SRCSSM
of Adelaide. We thank A.W. Thomas and A. Williams for their support and the
stimulating atmosphere of the SRCSSM.

\section{Appendice }

We study the pion propagation in the nuclear medium. In addition to the excitation
of particle-hole states by the s and p-wave couplings, the in-medium pion self-energy
\( S \) receives a contribution from the pion loop. For a pion of 4-momentum
\( q \) and isospin label \( a \) the one loop self energy writes:\begin{equation}
\label{A1}
S^{{\rm loop}}(q)\equiv S^{{\rm loop}}(\omega \, \, \vec{q})=\frac{1}{2}\int \frac{id^{4}k}{(2\pi )^{4}}\sum _{c}<q\, a;k\, c|{\cal M}|q\, a;k\, c>D_{R}(k),
\end{equation}
 where \( D_{R}(k) \) is the full in medium pion propagator and \( \cal M \)
is the (possibly in medium modified) \( \pi \, \pi  \) interaction which has
the decomposition: \textbf{}\begin{eqnarray}
<q_{1}\, a;k_{1}\, b|{\cal M}_{0}|q_{2}\, c;k_{2}\, ,d> & = & <q_{1};k_{1}|{\cal M}_{s}|q_{2};k_{2}>\delta _{ab}\delta _{cd}\nonumber \\
 & + & <q_{1};k_{1}|{\cal M}_{t}|q_{2};k_{2}>\delta _{ac}\delta _{bd}\nonumber \\
 & + & <q_{1};k_{1}|{\cal M}_{u}|q_{2};k_{2}>\delta _{ad}\delta _{bc},\label{A2} 
\end{eqnarray}
 and the projection on the total isospin states \( I=0,1,2 \) of the \( s \)
channel (\( s=(p+q)^{2} \) ) are:\begin{equation}
\label{A3}
{\cal M}_{0}=3{\cal M}_{s}+{\cal M}_{t}+{\cal M}_{u},\, \, \, \, {\cal M}_{1}={\cal M}_{t}-{\cal M}_{u},\, \, \, \, {\cal M}_{2}={\cal M}_{t}+{\cal M}_{u}.
\end{equation}
Working out the isospin summations one finds:\begin{equation}
\label{A4}
S^{{\rm loop}}(q)=\frac{1}{2}\int \frac{id^{4}k}{(2\pi )^{4}}\sum _{c}<q;k|\frac{1}{3}\left( {\cal M}_{0}+3{\cal M}_{1}+5{\cal M}_{2}\right) |q;k>D_{R}(k).
\end{equation}
 The particular combination \( \frac{1}{3}\left( {\cal M}_{0}+3{\cal M}_{1}+5{\cal M}_{2}\right) =3{\cal M}_{t}+{\cal M}_{s}+{\cal M}_{u} \)
is in fact the \( I=0 \) amplitude of the \( t \) channel: \begin{equation}
\label{A5}
S^{{\rm loop}}(q)=\frac{1}{2}\int \frac{id^{4}k}{(2\pi )^{4}}\sum _{c}<q;-q|{\cal M}_{0}|k;-k>D_{R}(k).
\end{equation}

Let us calculate, in the tree approximation, this \( I=0,\, \, \pi \pi  \)
amplitude, first ignoring possible in medium vertex corrections. The relevant
piece of the Lagrangian is: \begin{equation}
\label{A6}
{\cal L}_{\pi \pi }=\left( f_{\pi }+\theta \right) ^{2}\frac{1}{4}Tr\, \partial _{\mu }U\partial ^{\mu }U^{\dagger }+f_{\pi }m_{\pi }^{2}(f_{\pi }+\theta )\cos F\left( \frac{\phi }{f_{\pi }}\right) .
\end{equation}
 At the tree level we keep the terms of order \( \phi ^{4} \) and the \( \theta \pi \pi  \)
interaction term:\begin{eqnarray}
{\cal L}_{\pi \pi }^{(4)} & = & \frac{1}{f_{\pi }^{2}}\left[ -m_{\pi }^{2}\left( \alpha -\frac{1}{24}\right) \phi ^{4}+\left( \alpha -\frac{1}{6}\right) \phi ^{2}\partial _{\mu }\vec{\phi }.\partial ^{\mu }\vec{\phi }\right. \nonumber  \\
 &  & \left. +\left( 2\alpha +\frac{1}{6}\right) \vec{\phi }.\partial _{\mu }\vec{\phi }\, \vec{\phi }.\partial ^{\mu }\vec{\phi }\right] ,\label{A7a} \\
{\cal L}_{\pi \pi }^{\theta \pi \pi } & = & \frac{\theta }{f_{\pi }}\left( \partial _{\mu }\vec{\phi }.\partial ^{\mu }\vec{\phi }-\frac{1}{2}m_{\pi }^{2}\phi ^{2}\right) .\label{A7b} 
\end{eqnarray}
The \( {\cal M}_{s} \) amplitude is straightforwardly obtained as:\begin{eqnarray}
 &  & <q_{a};q_{b}|{\cal M}_{s}|q_{c};q_{d}>=\frac{1}{f_{\pi }^{2}}\left[ -\left( s-m_{\pi }^{2}\right) -2\left( \alpha -1/6\right) \sum _{i=a,\ldots d}\left( q_{i}^{2}-m_{\pi }^{2}\right) \right. \nonumber \\
 & + & \left. \left( s-q_{a}^{2}-q_{b}^{2}+m_{\pi }^{2}\right) \left( s-q_{c}^{2}-q_{d}^{2}+m_{\pi }^{2}\right) /s-m_{\theta }^{2}\right] \label{A8} 
\end{eqnarray}
 where \( s=(q_{a}+q_{b})^{2}=(q_{c}+q_{d})^{2} \) is the squared CM energy
of the pion pair. We see from Eq.(\ref{A8}) that in the low energy regime of
interest (\( q\sim m_{\pi }) \) the \( \theta  \) exchange contribution is
of order \( m_{\pi }^{2}/m_{\theta }^{2}. \) Since we limit ourselves to the
leading order in the chiral expansion we only keep the first contribution on
the RHS of Eq.(\ref{A8}), which is nothing but the well known non linear sigma
model result.

The \( {\cal M}_{t} \) and \( {\cal M}_{u} \) amplitudes are obtained by the
substitution \( (a\leftrightarrow c,\, \, s\leftrightarrow t) \) and \( (a\leftrightarrow d,\, \, s\leftrightarrow u) \)
respectively. It follows that the \( I=0 \) amplitude reads:\begin{equation}
\label{A9}
<q_{a};q_{b}|{\cal M}_{0}|q_{c};q_{d}>=\frac{1}{f_{\pi }^{2}}\left[ m_{\pi }^{2}-2s+\beta \sum _{i=a,\ldots d}\left( m_{\pi }^{2}-q_{i}^{2}\right) \right] ,\, \beta =1+10(\alpha -1/6),
\end{equation}
 where the relation \( s+t+u=\sum _{i=a,\ldots d}q_{i}^{2} \) has been used.
 
\begin{figure}
\epsfxsize=5cm
\centerline{\epsfbox{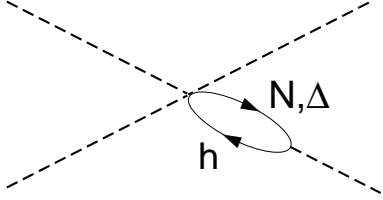}}
\caption{\label{fig-pi3}Vertex correction to \protect\( \pi \pi \protect \) scattering.}
\end{figure}

In the medium Chanfray and Davesne \cite{chanfray99a} have established that
the \( \pi \, \pi  \) interaction receives vertex corrections of the type shown
on {\LARGE }Fig. \ref{fig-pi3} The \( 3\pi \,  \)N vertex derived from the
Lagrangian Eq.(\ref{A21}) {\LARGE }is, at the relevant order:\begin{equation}
\label{A10}
{\cal L}_{3\pi N}=\frac{g_{A}}{2f_{\pi }^{3}}\overline{N}\gamma ^{\mu }\gamma _{5}\vec{\tau }N.\left[ \left( \alpha -1/6\right) \phi ^{2}\partial _{\mu }\vec{\phi }+(2\alpha +1/6)\vec{\phi }\, \vec{\phi }.\partial _{\mu }\vec{\phi }\right] .
\end{equation}
For a zero momentum pion pair \( (\vec{P}=0) \) the effective in medium \( I=0, \)
\( \pi \pi  \) potential takes the simple form:\begin{equation}
\label{A11}
<q_{a};q_{b}|{\cal M}_{0}^{{\rm eff}}|q_{c};q_{d}>=\frac{1}{f_{\pi }^{2}}\left[ m_{\pi }^{2}-2s+\beta \sum _{i=a,\ldots d}\left( m_{\pi }^{2}-q_{i}^{2}+\vec{q}_{i}^{\, 2}\tilde{\Pi }^{0}(\omega _{i},\vec{q}_{i})\right) \right] ,
\end{equation}
 where \( \vec{q}_{i}^{2}\tilde{\Pi }^{0}(\omega _{i},\vec{q}_{i}) \) is the
standard p-wave pionic polarisability which may include the screening effect
from short range correlations. Hence the effect of vertex corrections depending
on the p-wave polarisabilities is to make the effective \( \pi \, \pi  \) potential
independent of \( \alpha  \) for on shell quasi pions satisfying: \begin{equation}
\label{A12}
m_{\pi }^{2}-q_{i}^{2}+\vec{q}_{i}^{\, 2}\tilde{\Pi }^{0}(\omega _{i},\vec{q}_{i})=0.
\end{equation}
The pion loop contribution to the pion self energy Eq.(\ref{A5}) is obtained
from the matrix element of the \( I=0 \) amplitude making the replacements
\( q_{a}=q,\, q_{b}=-q,\, q_{c}=k,\, q_{d}=-k,\, s=t=0) \): \begin{equation}
\label{A13}
S^{{\rm loop}}(q)=\frac{1}{2}\int \frac{id^{4}k}{(2\pi )^{4}}\left[ m_{\pi }^{2}-2\beta \left( D_{R}^{-1}(k)+\omega ^{2}-\vec{q}^{\, 2}-m_{\pi }^{2}-\vec{q}^{\, 2}\tilde{\Pi }^{0}(\omega ,\vec{q})\right) \right] D_{R}(k).
\end{equation}
 Since we are interested in the effect of the in medium pion cloud we have to
substract the vacuum contribution. Constant terms such as \( D^{-1}D \) disappear.
One gets:\begin{eqnarray}
S^{{\rm loop}}(q) & = & \frac{1}{2}\int \frac{id^{4}k}{(2\pi )^{4}}\left[ m_{\pi }^{2}-2\beta \left( \omega ^{2}-\vec{q}^{\, 2}-m_{\pi }^{2}-\vec{q}^{\, 2}\tilde{\Pi }^{0}(\omega ,\vec{q})\right) \right] \left[ D_{R}(k)-D_{0}(k)\right] \nonumber  \\
 & = & \frac{<\phi ^{2}>}{6f_{\pi }^{2}}\left[ m_{\pi }^{2}-2\beta \left( \omega ^{2}-\vec{q}^{\, 2}-m_{\pi }^{2}-\vec{q}^{\, 2}\tilde{\Pi }^{0}(\omega ,\vec{q})\right) \right] ,\label{A14b} 
\end{eqnarray}
 which is valid to leading order in the pion density defined by:\begin{equation}
\label{A15}
<\phi ^{2}>=3\int \frac{id^{4}k}{(2\pi )^{4}}\left[ D_{R}(k)-D_{0}(k)\right] .
\end{equation}
 Finally the pion self energy has a contribution from the scalar field which
can be obtained directly from the \( \pi \pi \theta  \) Lagrangian:\begin{equation}
\label{A17}
S^{(\theta )}(q)=-(2q^{2}-m_{\pi }^{2})\frac{<\theta >}{f_{\pi }}.
\end{equation}
 The pion self energy is:\begin{eqnarray}
S(q) & = & S^{({\rm p-wave})}(q)+S^{({\rm loop})}(q)+S^{(\theta )}(q)\nonumber \\
 & = & \vec{q}^{\, 2}\tilde{\Pi }^{0}(\omega ,\vec{q})+\frac{<\phi ^{2}>}{6f_{\pi }^{2}}\left[ m_{\pi }^{2}-2\beta \left( \omega ^{2}-\vec{q}^{\, 2}-m_{\pi }^{2}-\vec{q}^{\, 2}\tilde{\Pi }^{0}(\omega ,\vec{q})\right) \right] \nonumber  \\
 &  & -(2q^{2}-m_{\pi }^{2})\frac{<\theta >}{f_{\pi }}.\label{A19} 
\end{eqnarray}

In the above expression we have ignored the s wave coupling. It would influence
the pion mass though the Born part of the \( \pi N \) amplitude. The pion propagator
at \( \vec{q}=0 \) writes:\begin{eqnarray}
\tilde{D}_{R}(\omega ) & = & \left[ \omega ^{2}-m_{\pi }^{2}-S(q)\right] ^{-1}=\frac{Z}{\omega ^{2}-m_{\pi }^{*2}},\label{A20} 
\end{eqnarray}
 with:\begin{equation}
\label{A21}
Z=1-\frac{\beta <\phi ^{2}>}{3f_{\pi }^{2}}-\frac{2<\theta >}{f_{\pi }^{2}},
\end{equation}
from which we deduce the effective pion mass, to lowest order in \( <\phi ^{2}> \)
and \( <\theta > \):\begin{eqnarray}
\left( \frac{m_{\pi }^{*}}{m_{\pi }}\right) ^{2} & = & Z\left( 1+(1+2\beta )\frac{<\phi ^{2}>}{6f_{\pi }^{2}}+\frac{<\theta >}{f_{\pi }}\right) \nonumber \\
 & = & \left( 1-\frac{<\theta >}{f_{\pi }}+\frac{<\phi ^{2}>}{6f_{\pi }^{2}}\right) .\label{A22} 
\end{eqnarray}
 As it should be the result is independent of \( \alpha , \) that is independent
of the choice of the canonical pion field.


\end{document}